\documentclass[showpacs,showkeys,nofootinbib,preprintnumbers,amsmath,amssymb]{revtex4}

\usepackage{graphicx}
\usepackage{dcolumn}
\usepackage{bm}

\def\bea{\begin{eqnarray}}
\def\eea{\end{eqnarray}}
\def\nnb{\nonumber}
\def\qqbar{\langle \bar{q}q\rangle}
\def\lamt{\tilde{\lambda}_\Lambda}

\begin{document}


\title{Scalar-Meson-Baryon Coupling Constants in QCD Sum Rules}

\author{G. Erkol}
\email{erkol@kvi.nl}
\author{R. G. E. Timmermans}
\email{timmermans@kvi.nl}
\affiliation{
          Theory Group, KVI,
          University of Groningen \\
          Groningen, The Netherlands}
\author{M. Oka}
\email{oka@th.phys.titech.ac.jp}
\affiliation{Department of Physics, H27, Tokyo Institute of Technology,
             Meguro, \\ Tokyo, 152-8551, Japan}          
\author{Th. A. Rijken}
\email{t.rijken@science.ru.nl}
\affiliation{
          Institute for Theoretical Physics,
          Radboud University Nijmegen \\
          Nijmegen, The Netherlands}

\date{\today}

\begin{abstract}
The external-field QCD Sum Rules method is used to evaluate the coupling
constants of the light-isoscalar scalar meson (``$\sigma$'' or $\epsilon$)
to the $\Lambda$, $\Sigma$, and $\Xi$ baryons. It is shown that these coupling
constants as calculated from QCD Sum Rules are consistent with $SU(3)$-flavor
relations, which leads to a determination of the $F/(F+D)$ ratio of the scalar
octet assuming ideal mixing: we find $\alpha_s \equiv F/(F+D)=0.55$. The
coupling constants with $SU(3)$ breaking effects are also discussed.   
\end{abstract}

\maketitle

\section{Introduction}
Hadronic interactions are in principle explained by quantum chromodynamics
(QCD). Such a first-principles description of the hadron-hadron interaction,
however, is highly complicated, particularly at low energy, where QCD is a
nonperturbative theory. In practice, therefore, effective hadronic Lagrangians
are often used. The coupling constants at the hadronic vertices are then among
the most fundamental quantities that should be computed from QCD.

There is a long history of successful approaches of describing the two-baryon
interaction using meson-exchange potentials. The values of the meson-baryon
coupling constants have been empirically  
determined so as to reproduce the nucleon-nucleon (N\!N)~\cite{Nag78,Sto94}, hyperon-nucleon (YN)~\cite{Nag78x,Mae89,Rij98} and hyperon-hyperon (YY) interactions in terms of {\em e.g.} one-boson exchange (OBE) models. The scalar mesons play significant roles in such phenomenological potential models. In early OBE models for the NN interaction the exchange of an isoscalar-scalar ``$\sigma$'' meson with a mass of about 500 MeV was needed to obtain enough medium-range attraction and a sufficiently strong spin-orbit force. It was only later understood that the exchange of a broad isoscalar-scalar meson, the $\varepsilon$(760)~\cite{Pro73,Sve92}, simulates the exchange of such a low-mass ``$\sigma$''~\cite{Bin72}. 

Existence of the ``$\sigma$'' meson is expected also from chiral symmetry of QCD. The $\sigma$ meson appears as a chiral partner of the Nambu-Goldstone boson, $\pi$, and thus plays the role of the ``Higgs boson'' in chiral symmetry breaking. Recent analyses of $\pi-\pi$ scattering have revealed a broad  
resonance at the mass around 600 MeV, called $f_0(600)$ by the Particle Data Group~\cite{Eid04}, which is considered to play the role of $\varepsilon$(760) in the boson exchange potential of the baryonic interactions. 

In terms of OBE models for two-baryon interactions, the dominant contribution to $\Lambda\Lambda$ interaction comes from the scalar $\varepsilon(760)$ meson exchange~\cite{Afn03, Fer05}. The recent identification of $^6_{\Lambda\Lambda}He$ and the measurement of the $\Lambda\Lambda$ pair suggest that the binding energy of $\Lambda\Lambda$ ($\Delta B_{\Lambda\Lambda}\simeq 1.0\,$ MeV) is considerably smaller than the binding energy of $N\!N$~\cite{Tak01}. This is in contrast to the outcome of the earlier measurement which is $\Delta B_{\Lambda\Lambda}\simeq 4.7\,$ MeV~\cite{Pro66}. This issue has also been examined within the framework of Nijmegen OBE potential model D (NHC-D) \cite{Nag76}. In this model, the pseudoscalar octet {$\pi$, $\eta$, $\eta^\prime$, $K$}, the vector octet {$\rho$, $\phi$, $\omega$, $K^\ast$} and the scalar singlet {$\epsilon$} are the exchanged mesons and the coupling constants are fitted to data while other physical properties of the particles are taken from experiment. The estimated value of $\Delta B_{\Lambda\Lambda}$ in this model implies a rather strong attractive $\Lambda\Lambda$. However, in the Nijmegen soft-core (NSC) potential models~\cite{Mae89, Rij98}, where there is a scalar nonet instead of a scalar singlet, we have much weaker attractive potentials than in the case of NHC-D in the $\Lambda\Lambda$ systems. In this framework, since the scalar $\varepsilon(760)$ exchange plays the most crucial role in $\Lambda\Lambda$ interactions, it is necessary to determine the $\Lambda\Lambda\varepsilon$ coupling constant in a model independent way in order to understand the role of $\varepsilon$ exchange in the strangeness $S=-2$ sector.  

The structure and even the status of the scalar mesons, however, have always been controversial~\cite{Tim94,Swa94}. In the quark model, the simplest assumption for the structure of the scalar mesons is the $^3P_0$ $q\bar{q}$ states. In this case, the scalar mesons might form a complete nonet of dressed $q\bar{q}$ states, resulting from {\em e.g.} the coupling of the $P$-wave $q\bar{q}$ states to meson-meson channels~\cite{Bev86}. Explicitly, the unitary singlet and octet states, denoted respectively by $\varepsilon_1$ and $\varepsilon_8$, read
\bea\label{octsing}
   \varepsilon_1 & = & (u\bar{u}+d\bar{d}+s\bar{s})/\sqrt{3} \, , \nnb \\
   \varepsilon_8 & = & (u\bar{u}+d\bar{d}-2s\bar{s})/\sqrt{6}\, .
\eea 
The physical states are mixtures of the pure $SU(3)$-flavor states,
and are written as
\bea\label{mix}
   \varepsilon & = & \cos\theta_s\,\varepsilon_1+\sin\theta_s\,\varepsilon_8
                     \, , \nnb \\
           f_0 & = &-\sin\theta_s\,\varepsilon_1+\cos\theta_s\,\varepsilon_8 
                     \, .
\eea
For ideal mixing holds that $\tan\theta_s=1/\sqrt{2}$ or $\theta_s\simeq 35.3^\circ$, and thus one would identify \bea\label{psqqb}
   \varepsilon(760) & = & (u\bar{u}+d\bar{d})/\sqrt{2}\, , \nnb \\
           f_0(980) & = & -s\bar{s} \, .
\eea
The isotriplet member of the octet is $a_0^{\pm,0}$(980), where 
\bea\label{azero}
   a_0^0(980) = (u\bar{u}-d\bar{d})/\sqrt{2} \, .
\eea

An alternative and arguably more natural explanation for the masses and decay properties of the
lightest scalar mesons is to regard these as cryptoexotic $q^2\bar{q}^2$
states~\cite{Jaf77}. In the MIT bag model, the scalar $q\bar{q}$ states are
predicted around $1250$ MeV, while the attractive color-magnetic force results
in a low-lying nonet of scalar $q^2\bar{q}^2$ mesons~\cite{Jaf77,Aer80}.
This nonet contains a nearly degenerate set of $I=0$ and $I=1$ states, which
are identified as the $f_0(980)$ and $a^{\pm,0}_0(980)$ at the $\bar{K}K$
threshold, where
\bea\label{ps4q}
   a^0_0(980) & = & (sd\bar{s}\bar{d}-su\bar{s}\bar{u})/\sqrt{2} \, , \nnb \\
     f_0(980) & = & (sd\bar{s}\bar{d}+su\bar{s}\bar{u})/\sqrt{2} \, ,
\eea
with the ideal-mixing angle $\tan\theta_s=-\sqrt{2}$ or $\theta_s\simeq
 -54.8^\circ$ in this case. The light isoscalar member of the nonet is
\bea\label{sigmaq4}
   \varepsilon(760) & = & ud\bar{u}\bar{d} \, .
\eea
The nonet is completed by the strange member $\kappa$(880), which like the
$\varepsilon$(760) is difficult to detect because it is hidden under the
strong signal from the $K^*$(892)~\cite{Tim94,Swa94}. We shall use in this paper the
nomenclature $(a_0^{\pm,0},f_0,\sigma,\kappa)$ for the scalar-meson nonet,
where one should identify $\sigma=\varepsilon(760)$. 

One way to make progress with the scalar mesons is to study their role in the various two-baryon reactions (NN, YN, YY). Our aim in this paper is to calculate the $\Lambda\Lambda\sigma$, $\Xi\Xi\sigma$ and $\Sigma\Sigma\sigma$ coupling constants, using the QCD Sum Rules (QCDSR) method. The QCDSR method~\cite{Shi79} is a powerful tool to extract qualitative and quantitative information about hadron properties~\cite{Rei84, Col00}. In this framework, one starts with a correlation function, which is constructed in terms of hadron interpolating fields. On the theoretical side, the correlation function is calculated using the Operator Product Expansion (OPE) in the Euclidian region. This correlation function is matched with an \emph{Ansatz} which is introduced from the hadronic degrees of freedom on the phenomenological side, and this matching provides a determination of the hadronic parameters like baryon masses, magnetic moments, coupling constants of hadrons and so on.

There are different approaches in constructing the QCDSR (see {\em e.g.}~Ref.~\cite{Col00} for a review). One usually starts with the vacuum-to-vacuum matrix element of the correlation function that
is constructed with the interpolating fields of two baryons and one meson. However, this three-point function method has as a major drawback that at low momentum transfer the OPE fails. Moreover, when the momentum of the meson is large, it is plagued by problems with higher resonance contamination~\cite{Mal97}. The other method that is free from the above problems is the external-field
method. There are two formulations that can be used to construct the external-field sum rules: In the vacuum-to-meson method, one starts with a vacuum-to-meson transition matrix element of the baryon interpolating fields, where some other transition matrix elements should be evaluated~\cite{Rei84}. (This is also the starting point of the light-cone QCDSR method.) In Ref.~\cite{Shi95},  pion-nucleon coupling constant was calculated in the soft meson limit using this approach. Later it was pointed out that the sum rule for pion-nucleon coupling in the soft-meson limit can be reduced to the sum rule for the nucleon mass by a chiral rotation so the coupling was calculated again with a finite meson momentum~\cite{Bir96}. These calculations were improved considering the coupling schemes at different Dirac structures and beyond the chiral limit contributions~\cite{Lee98,Kim98,Kim99}. In this paper, we calculate the baryon-sigma meson coupling constants, using the external field QCDSR method \cite{Iof84}. We evaluate the vacuum to vacuum transition matrix element of two baryon interpolating fields in an external sigma field and construct the sum rules. This method has been used to determine the magnetic moments of baryons~\cite{Iof84,Bal83,Chi86,Chi85}, the nucleon axial coupling constant~\cite{Chi85,Bel84}, the nucleon sigma term~\cite{Jin93}, and baryon isospin mass splittings~\cite{Jin95}. It has also been shown that at low momentum transfer, this method is very successful in evaluating the hadronic coupling constants. Recently, the $N\!N\sigma$ coupling constant, $g_{N\!N\sigma}$, was calculated using this method~\cite{Erk05}. It has also been applied, previously, to the calculations of the strong and weak parity violating pion-nucleon coupling constants~\cite{Hwa96,Hwa97,Hen96} and the coupling constants of the vector mesons $\rho$ and $\omega$ to the nucleon~\cite{Wen97}.

In the $SU(3)$ flavor symmetric one can classify the meson-baryon coupling constants in terms of two parameters, the $N\!Na_0$ coupling constant, $g_{N\!Na_0}$ and the $F/(F+D)$ ratio of the scalar octet, $\alpha_s$~\cite{Swa63}:
\bea\label{relSU3} g_{N\!Na_0}&=&g\,,~~~~~~ g_{N\!N\varepsilon_8}=\frac{1}{\sqrt{3}}\,g\,(4\alpha_s-1)\,,~~~~~~ g_{\Lambda\Lambda\varepsilon_8}=-\frac{2}{\sqrt{3}}\,g\,(1-\alpha_s)\,,\\ g_{\Xi\Xi\varepsilon_8}&=&-\frac{1}{\sqrt{3}}\,g\,(1+2\alpha_s)\,,~~~~~~ g_{\Sigma\Sigma\varepsilon_8}=\frac{2}{\sqrt{3}}\,g\,(1-\alpha_s)\,,~~~~~~g_{\Xi\Xi a_0}=g\,(2\alpha_s-1)\,,\nnb\\ g_{\Sigma\Sigma a_0}&=&2\,g\,\alpha_s\,, ~~~~~~ g_{\Lambda\Sigma a_0}=\frac{2}{\sqrt{3}}\,g\,(1-\alpha_s)\,,~~~~~~ g_{\Lambda N\kappa}=-\frac{1}{\sqrt{3}}\,g\,(1+2\alpha_s)\,,\nnb\\ g_{\Sigma N \kappa}&=&g\,(1-2\alpha_s)\,.\nnb\eea
Considering the mixing between the singlet and the octet members of the scalar nonet, one obtains for $g_{B\!B\sigma}$ and $g_{B\!Bf_0}$,
\bea\label{Noctsing}
g_{B\!B\sigma} &=& \cos~\theta_s g_1 + \sin~\theta_s g_{B\!B\varepsilon_8}\,,\nnb\\
g_{B\!B f_0} &=& -\sin~\theta_s g_1 + \cos~\theta_s g_{B\!B\varepsilon_8}\,,
\eea
where $g_1=g_{B\!B\varepsilon_1}$ is the flavor singlet coupling, and $\theta_s$ is the scalar mixing angle.

We shall first consider the sum rules in the $SU(3)$ flavor symmetric limit to see if the predicted values for the meson baryon coupling constants from the sum rules are consistent with the $SU(3)$ relations. We show that this is indeed the case which leads to a determination of the $F/(F+D)$ ratio of the scalar octet. Furthermore, keeping track of these coupling constants with the $SU(3)$ relations, we obtain the values of the other scalar meson-baryon coupling constants. For this purpose, we assume ideal mixing and make the analysis in both $q\bar{q}$ and $q^2\bar{q}^2$ pictures for the scalar mesons. As we move from the $S=0$ to the $S=-1$ and $S=-2$ sectors, the flavor $SU(3)$ breaking occurs as a result of the $s$-quark mass and the physical masses of the baryons and mesons. We also consider the $SU(3)$ breaking effects for the sum rules to estimate the amount of breaking, individually for each coupling.

We have organized our paper as follows: in Section~\ref{secNSR}, we present
the formulation of QCDSR with an external scalar field and construct the sum
rules for the $\Lambda\Lambda\sigma$, $\Xi\Xi\sigma$ and $\Sigma\Sigma\sigma$
coupling constants. We give the numerical analysis and
discuss the results in Section~\ref{secAN}. Finally, we arrive at our conclusions in Section~\ref{secCONC}.

\section{Baryon Sum Rules in an external sigma field}\label{secNSR}
\subsection{Construction of the Sum Rules}
In the external-field QCDSR method one starts with the correlation function of the baryon interpolating fields in the presence of an external constant isoscalar-scalar field $\sigma$, defined by the following:
\bea \label{cor1}\Pi_{B\sigma}(q)=i\int d^4 x~ e^{i q\cdot x}\, \Big
\langle 0\Big |{\cal T}[\eta_B(x)\bar{\eta}_B(0)]\Big |0\Big
\rangle_\sigma \,\, , \eea where $\eta_B$ are the baryon interpolating fields which are chosen as~\cite{Rei84}
 \bea\label{intfi}
\eta_\Xi&=&\epsilon_{abc}[(s_a^T C \gamma_\mu s_b)\gamma_5\gamma^\mu u_c]\,,\\
\eta_\Sigma&=&\epsilon_{abc}[(u_a^T C \gamma_\mu u_b)\gamma_5\gamma^\mu s_c]\,,\nnb\\
\eta_\Lambda&=&(2/3)^{1/2}\epsilon_{abc}[(u_a^T C \gamma_\mu s_b)\gamma_5\gamma^\mu d_c-(d_a^T C \gamma_\mu s_b)\gamma_5\gamma^\mu
u_c]\,\,\,\nnb. \eea for $\Xi$, $\Sigma$ and $\Lambda$, respectively. Here $a, b, c$ denote the color indices, and $T$ and $C$ denote transposition and charge conjugation, respectively. For the interpolating field of each octet baryon, there are two independent local operators, but the ones in Eq.~(\ref{intfi}) are the optimum choices for the lowest-lying positive parity baryons (see {\em e.g.} Ref~\cite{Jid96} for a discussion on negative-parity baryons in QCDSR). 

The external sigma field contributes the correlation function in Eq.~(\ref{cor1}) in two ways: first, it directly couples the quark field in the baryon current and second, it modifies the condensates by polarizing the QCD vacuum. In the presence of the external scalar field there are no correlators that break the Lorentz invariance; however, the correlators already existing in the vacuum are modified by the external field: 
\bea\label{vaccon}\qqbar_\sigma &\equiv& \qqbar
+ g^\sigma_q \chi \sigma \qqbar\,, \\\langle g_c \bar{q}
{\bm\sigma}\cdot {\bm G} q\rangle_\sigma &\equiv& \langle g_c
\bar{q} {\bm\sigma}\cdot {\bm G} q\rangle + g^\sigma_q \chi_G
\sigma \langle g_c \bar{q} {\bm\sigma}\cdot {\bm G} q\rangle\,
,\nnb\eea where only the responses linear in the external-field are taken into account. Here, $g^\sigma_q$ is the quark-$\sigma$ coupling constant and $\chi$ and $\chi_G$ are the susceptibilities corresponding to quark and quark-gluon mixed condensates, respectively. In Eq.~(\ref{vaccon}), $\langle\bar{q}q\rangle$ represents either $\langle\bar{u}u\rangle$ or $\langle\bar{d}d\rangle$, as we have assumed that $\langle\bar{u}u\rangle \simeq \langle\bar{d}d\rangle$ and the responses of the up and the down quarks to the external isoscalar field are the same. Note that, here we assume ideal mixing in the scalar sector, that is, we take the sigma meson without a strange-quark content. Therefore, the sigma meson couples only to the $u$- or the $d$-quark in the baryon, where we take $g_u^\sigma=g_d^\sigma$ and $g_s^\sigma=0$.

In the Euclidian region, the OPE of the product of two interpolating fields can be written as follows: \bea
\label{opex}\Pi_{B\sigma}(q)=\sum_{n} C^\sigma_n(q) O_n \,\, ,\eea where $C^\sigma_n(q)$ are the Wilson coefficients and $O_n$ are the local operators in terms of quarks and gluons.  In order to calculate the Wilson coefficients, we need the quark propagator in the presence of the external sigma field. In coordinate space the full quark propagator takes the form: \bea\label{proptot} S_q(x)=S_q^{(0)}(x)+S_q^{(\sigma)}(x)\, ,\eea where,

\bea\label{prop0} i~S_q^{(0)ab}&\equiv&\langle 0|T[q^a(x)
\bar{q}^b(0)|0\rangle_0\\\nnb\\ &=&
\frac{i~\delta^{ab}}{2\pi^2x^4}\hat{x}-\frac{i~\lambda_{ab}^n}{32
\pi^2}\frac{g_c}{2} G_{\mu\nu}^n
\frac{1}{x^2}(\sigma^{\mu\nu}\hat{x}+\hat{x}
\sigma^{\mu\nu})-\frac{\delta^{ab}}{12}\langle\bar{q}q\rangle-\frac{\delta^{ab}x^2}{192}\langle
g_c \bar{q}{\bm\sigma}\cdot {\bm G} q\rangle
\nnb\\\nnb\\&&-\frac{m_q \delta^{ab}}{4\pi^2
x^2}-\frac{m_q}{32\pi^2}\lambda_{ab}^n g_c
G_{\mu\nu}^n\sigma^{\mu\nu} \ln(-x^2)-\frac{\delta^{ab}\langle
g_c^2 G^2\rangle}{2^9\times 3 \pi^2} m_q x^2
\ln(-x^2)\nnb\\\nnb\\&&+\frac{i~\delta^{ab}m_q}{48}\qqbar\hat{x}+\frac{i~m_q\delta^{ab}x^2}{2^7\times
3^2}\langle g_c \bar{q}{\bm\sigma} \cdot {\bm G} q\rangle\hat{x}+O(\alpha_s^2\,,\,m_q^2)
\, ,\nnb\eea and

\bea\label{propE} i~S_q^{(\sigma)ab} &\equiv&\langle 0|T[q^a(x)
\bar{q}^b(0)|0\rangle_\sigma\\\nnb\\&=&g^\sigma_q \sigma
\Big[-\frac{ \delta^{ab}}{4\pi^2
x^2}-\frac{1}{32\pi^2}\lambda_{ab}^n g_c
G_{\mu\nu}^n\sigma^{\mu\nu} \ln(-x^2)-\frac{\delta^{ab}\langle
g_c^2 G^2\rangle}{2^9\times 3 \pi^2} x^2
\ln(-x^2)\nnb\\\nnb\\&&~~~~~~+\frac{i~\delta^{ab}}{48}\qqbar\hat{x}-\frac{\delta^{ab}\chi}{12}\qqbar+
\frac{i~\delta^{ab}x^2}{2^7\times 3^2}\langle g_c
\bar{q}{\bm\sigma} \cdot {\bm G}
q\rangle\hat{x}\nnb\\\nnb\\&&~~~~~~ -\frac{\delta^{ab}x^2}{192}
\chi_G \langle g_c \bar{q}{\bm\sigma}\cdot {\bm G} q\rangle\Big
]\,+O(\sigma^2)\, .\nnb\eea Here, $G^{\mu\nu}$ is the gluon tensor and $g_c^2=4\pi\alpha_s$ is the quark-gluon coupling constant squared. Note that, in the quark propagator above, we have included the terms that are proportional to the quark masses, $m_q$, since these terms give non-negligible contributions to the final result as far as the strange quark mass is considered.

Using the quark propagator in Eq.~(\ref{proptot}), one can compute the correlation function $\Pi_{B\sigma}(q)$. The Lorentz covariance and parity implies the following form for $\Pi_{B\sigma}(q)$: \bea\Pi_{B\sigma}(q)=(\Pi_{B0}^1+\Pi_{B0}^q~ \hat{q})+(\Pi_{B\sigma}^1+\Pi_{B\sigma}^q~ \hat{q})\sigma+O(\sigma^2)\,,\eea where $\hat{q}=q^\mu \gamma_\mu$ is the four momentum of the baryon. Here $\Pi_{B0}^1$ and $\Pi_{B0}^q$ represent the invariant functions in the vicinity of the external field, which can be used to construct the mass sum rules for the relevant baryons, and $\Pi_{B\sigma}^1$ and $\Pi_{B\sigma}^q$ denote the invariant functions in the presence of the external field. Using these invariant functions, one can derive the sum rules at the structures $1$ and $\hat{q}$. In Ref.~\cite{Erk05} it was found that the sum rule at the structure $\hat{q}$ for the $N\!N\sigma$ coupling constant is more stable than the other sum rule at the structure $1$, with respect to variations in the Borel mass. Motivated with this, we here present only the sum rules at the structure $\hat{q}$ and use these for the determination of the coupling constants. 

\subsection{$\Lambda$ Sum Rules and $\Lambda\Lambda\sigma$ Coupling Constant}
We shall first present the sum rules calculations for $\Lambda\Lambda\sigma$ coupling constant in detail and in the next subsection, we shall give the sum rules for $\Xi\Xi\sigma$ and $\Sigma\Sigma\sigma$ couplings. At the quark level, we have for $\Lambda$:
\bea\label{cor2}\Big \langle 0\Big |{\cal
T}[\eta_\Lambda(x)\bar{\eta}_\Lambda(0)]\Big |0\Big \rangle_\sigma= \frac{2i}{3}
\epsilon^{abc}\epsilon^{a^\prime b^\prime c^\prime} &&\Big(Tr \{S_u^{a
a^\prime}(x) \gamma_\nu C [S_s^{b b^\prime}(x)]^T C
\gamma_\mu\}\gamma_5 \gamma^\mu S_d^{c
c^\prime}(x)\gamma^\nu\gamma_5\,\nnb\\&&+ Tr \{S_d^{c
c^\prime}(x) \gamma_\nu C [S_s^{b b^\prime}(x)]^T C
\gamma_\mu\}\gamma_5 \gamma^\mu S_u^{a
a^\prime}(x)\gamma^\nu\gamma_5\,\nnb\\ &&-\gamma_5 \gamma_\mu S_d^{c
c^\prime}(x) \gamma_\nu C [S_s^{b b^\prime}(x)]^T C \gamma^\mu S_u^{a
a^\prime}(x) \gamma^\nu \gamma_5\,\nnb\\ &&-\gamma_5 \gamma_\mu S_u^{a
a^\prime}(x) \gamma_\nu C [S_s^{b b^\prime}(x)]^T C \gamma^\mu S_d^{c
c^\prime}(x) \gamma^\nu \gamma_5\Big) . \eea

Using the quark propagator in Eq.~(\ref{proptot}), the invariant function at the structure $\hat{q}$ in the presence of the external field, $\Pi_{\Lambda\sigma}^q$, is calculated as:

\bea\label{qmom} \Pi^q_{\Lambda\sigma}(q)= g_q^\sigma\, \frac{1}{(2\pi)^4}\,&&\Bigg[\frac{4}{3}a_q\,(1-f)\,\ln(-q^2)
-\frac{4}{9q^2}\chi\,a_q^2(1+2f) -(\chi +\chi_G)\frac{m_0^2}{18q^4}\,a_q^2(1+2f)\nnb\\&&+ \frac{5}{12q^2}m_0^2\, a_q-\frac{2}{3}(2m_s-m_q)q^2\,\ln(-q^2)-\frac{2}{3}\chi\,  a_q\,(2m_s-m_q)\ln(-q^2) \nnb\\&&+\frac{1}{9q^4}\Big[(5 m_s-3 m_q) + 8(m_s-m_q)f\Big]\,a_q^2\Bigg],\eea where we have defined $f=\frac{\langle\bar{s}s\rangle}{\langle\bar{q}q\rangle}-1$, $a_q=-(2\pi)^2 \langle\bar{q}q\rangle$ and $\langle g_c\bar{q} {\bm\sigma} \cdot {\bm G} q\rangle = m_0^2 \langle\bar{q}q\rangle$. 

In order to construct the hadronic side, we saturate the correlator in Eq.(\ref{cor1}) with $\Lambda$ states and write, \bea\label{sat}
\Pi_{\Lambda\sigma}(q)=\frac{\langle 0 | \eta_\Lambda | \Lambda \rangle}{q^2-M_\Lambda^2}\,\,
\langle \Lambda|\sigma \Lambda\rangle \,\,\frac{\langle \Lambda| \bar{\eta}_\Lambda | 0
\rangle}{q^2-M_\Lambda^2}\,\,\,, \eea where $M_\Lambda$ is the mass of the
$\Lambda$. The matrix element of the current $\eta_\Lambda$ between the
vacuum and the $\Lambda$ state is defined as, \bea\label{overlap}
\langle 0 | \eta_\Lambda | \Lambda \rangle= \lambda_\Lambda \upsilon\,\,\,, \eea
where $\lambda_\Lambda$ is the overlap amplitude and $\upsilon$ is the Dirac
spinor for the $\Lambda$, which is normalized as
$\bar{\upsilon}\upsilon=2M_\Lambda$. Inserting Eq.~(\ref{overlap}) into
Eq.~(\ref{sat}) and making use of the isoscalar scalar
meson-baryon interaction Lagrangian density \bea {\cal L}=-g_{\Lambda \Lambda \sigma}\,
\bar{\upsilon}\upsilon~\sigma \,\,\, , \eea we obtain the hadronic part as:
\bea\label{phpart}-|\lambda_\Lambda|^2\frac{\hat{q}+M_\Lambda}{q^2-M_\Lambda^2} g_{\Lambda\Lambda \sigma}\frac{\hat{q}+M_\Lambda}{q^2-M_\Lambda^2}\,\,\,.\eea

We have also contributions coming from the excitations to higher $\Lambda$ states which are written as, \bea\label{phpartex}-\lambda_\Lambda \lambda_{\Lambda^\ast}\,\frac{\hat{q}+M_\Lambda}{q^2-M_\Lambda^2}\, g_{\Lambda\Lambda^\ast \sigma}\,\frac{\hat{q}+M_{\Lambda^\ast}}{q^2-M_{\Lambda^\ast}^2}\,\,\,,\eea and the ones coming from the intermediate states due to $\sigma$-$\Lambda$ scattering i.e. the {\it continuum} contributions. Note that the term that corresponds to the excitations to higher $\Lambda$ states also has a pole at the $\Lambda$ mass, but a single pole instead of a double one like in Eq.~(\ref{phpart}). This single pole term is not damped after the Borel transformation and should be included in the calculations. There is another contribution that comes from the response of the continuum to the external field, which is given as: \bea\label{cont}\int^\infty_0 \frac{-\Delta s_0 ~b(s)}{s-q^2} \delta(s-s_0) ds\, ,\eea where $s_0$ is the continuum threshold, $\Delta s_0$ is the response of the continuum threshold to the external field and $b(s)$ is a function that is calculated from OPE. When $\Delta s_0$ is large, this term should also be included in the hadronic part \cite{Iof95}.

Matching the OPE side with the hadronic side and applying the Borel transformation, the sum rule for $\Lambda\Lambda\sigma$ coupling at the structure $\hat{q}$ is obtained as:
\bea \label{sumq} & \Bigg\{&-\frac{4}{3}\,M^4\,a_q\,(1-f)\,E^\Lambda_0+\frac{4M^2}{9}\chi\,a_q^2(1+2f)\,L^{4/9}-\frac{5}{12}\, M^2\, m_0^2\, a_q\,L^{-14/27}\\
&-&(\chi +\chi_G)\frac{m_0^2}{18}\,a_q^2(1+2f)\,L^{-2/27}
+\frac{2}{3}(2m_s-m_q)M^6\,E^\Lambda_1\,L^{-8/9}\nnb\\&+&\frac{2}{3}\chi\,  
a_q\,(2 m_s-m_q)\,M^4\, E^\Lambda_0\,L^{-4/9}
+\frac{1}{9}\Big[(5 m_s-3 m_q) + 8(m_s-m_q)f\Big]\,a_q^2 \Bigg\}~e^{M_\Lambda^2/M^2}\nnb\\
& &~~~~~~~~=-\lamt^2\, \frac{M_\Lambda}{g_q^\sigma}\, g_{\Lambda\Lambda\sigma} + \tilde{B}_\Lambda\, \frac{M^2}{g_q^\sigma}+\frac{\Delta s^\Lambda_0}{2 g_q^\sigma}\Big[(s^\Lambda_0)^2-4\,m_s\,f\,a_q\Big] \, M^2\,L^{-4/9} e^{(M_\Lambda^2-s_0^\Lambda)/M^2}\, ,\nnb\eea
where we have defined $\lamt^2=32 \pi^4 \lambda_\Lambda^2$ and $M$ is the Borel mass.
The continuum contributions are included by the factors \bea
E^\Lambda_0&\equiv& 1- e^{-s_0^\Lambda/M^2}\, ,\nnb\\E^\Lambda_1&\equiv& 1-
e^{-s_0^\Lambda/M^2}\Big(1+\frac{s_0^\Lambda}{M^2}\Big)\, ,
\eea where $s_0^\Lambda$ is the continuum threshold. In the sum rule above, we have included the single pole contribution with the factors $\tilde{B}_\Lambda$. The third term on the right
hand side (RHS) of Eq.~(\ref{sumq}) denotes the contribution that is explained in Eq.~(\ref{cont}). Note that this term is suppressed by the factor $e^{-(s_0^i-M_\Lambda^2)/M^2}$
as compared to the single pole term. We have incorporated the effects of the anomalous dimensions of various operators through the factor $L=\ln(M^2/\Lambda_{QCD}^2)/\ln(\mu^2/\Lambda_{QCD}^2)$, where $\mu$ is the renormalization scale and $\Lambda_{QCD}$ is the QCD scale parameter.

\subsection{$\Xi$ and $\Sigma$ Sum Rules and $\Xi\Xi\sigma$ and $\Sigma\Sigma\sigma$ Coupling Constants}
One can apply the method explained in the previous subsection for the $\Xi\Xi\sigma$ and $\Sigma\Sigma\sigma$ couplings and derive the corresponding sum rules. Using the interpolating fields in Eq.~(\ref{intfi}), we obtain
\bea\label{cor2sigcas}\Big \langle 0\Big |{\cal
T}[\eta_\Xi(x)\bar{\eta}_\Xi(0)]\Big |0\Big \rangle_\sigma&=& 2i
\epsilon^{abc}\epsilon^{a^\prime b^\prime c^\prime} Tr \{S_s^{a
a^\prime}(x) \gamma_\nu C [S_s^{b b^\prime}(x)]^T C
\gamma_\mu\}\gamma_5 \gamma^\mu S_u^{c
c^\prime}(x)\gamma^\nu\gamma_5\,,\\ \Big \langle 0\Big |{\cal
T}[\eta_\Sigma(x)\bar{\eta}_\Sigma(0)]\Big |0\Big \rangle_\sigma&=& 2i
\epsilon^{abc}\epsilon^{a^\prime b^\prime c^\prime} Tr \{S_u^{a
a^\prime}(x) \gamma_\nu C [S_u^{b b^\prime}(x)]^T C
\gamma_\mu\}\gamma_5 \gamma^\mu S_s^{c
c^\prime}(x)\gamma^\nu\gamma_5\nnb\,,\eea
at the quark level for $\Xi$ and $\Sigma$, respectively. Using the quark propagator in Eq.~(\ref{proptot}), the invariant functions at the structure $\hat{q}$ are calculated as:
\bea\label{Cqmom} \Pi^q_{\Xi\sigma}(q)= g_q^\sigma\, \frac{1}{(2\pi)^4}\,\Big[\frac{m_0^2}{6\,q^2}\,a_q+a_q\,\ln(-q^2)\Big]\,,\eea

\bea\label{Sqmom} \Pi^q_{\Sigma\sigma}(q)= g_q^\sigma\, \frac{1}{(2\pi)^4}&&\Big[\frac{m_0^2}{3\,q^2}\,a_q-(\chi_g+\chi)\frac{m_0^2}{6\,q^4}\,a_q^2 -\frac{4\chi}{3\,q^2}\,a_q^2-2m_q\,q^2\,\ln(-q^2)\nnb\\&&-2m_q\,\chi\,a_q\,\ln(-q^2)+\frac{2m_q}{3q^4}\,a_q^2\Big]\,.\eea
The sum rules are obtained by matching the OPE side with the hadronic side and applying the Borel transformation. As a result of this operation, we obtain:
\bea\label{Csumq}
& \Big[&-\frac{m_0^2}{6}a_q\,M^2\,L^{-14/27}-a_q\,M^4\,E^\Xi_0 \Big]~e^{M_\Xi^2/M^2}\nnb\\
& &~~~~~~~~=-\tilde{\lambda}_\Xi^2\, \frac{M_\Xi}{g_q^\sigma}\, g_{\Xi\Xi\sigma} + \tilde{B}_\Xi\, \frac{M^2}{g_q^\sigma}+\frac{(s^\Xi_0)^2}{2  
g_\Xi^\sigma} \,\Delta s^\Xi_0\, M^2\,L^{-4/9} e^{(M_\Xi^2-s_0^\Xi)/M^2}\, ,\\\nnb
\eea
and
\bea\label{Ssumq}
& \Big[&-\frac{m_0^2}{3}\,a_q\,M^2\,L^{-14/27}-(\chi_G+\chi)\frac{m_0^2}{6}\,a_q^2\,L^{-2/27}+\frac{4}{3}\,\chi\,a_q^2\,M^2\,L^{4/9}+2\,m_q\,M^6\,E^\Sigma_1\,L^{-8/9}\nnb\\&&+2\,\chi\,m_q\,a_q\,M^4\,E^\Sigma_0\,L^{-4/9}+\frac{2}{3}m_q\,a_q^2 \Big]~e^{M_\Sigma^2/M^2}\\
& &~~~~=-\tilde{\lambda}_\Sigma^2\, \frac{M_\Sigma}{g_q^\sigma}\, g_{\Sigma\Sigma\sigma} + \tilde{B}_\Sigma\, \frac{M^2}{g_q^\sigma}+\frac{\Delta s^\Sigma_0}{2 g_q^\sigma}[(s^\Sigma_0)^2-2\,m_s(f+1)a_q] \, M^2\,L^{-4/9} e^{(M_\Sigma^2-s_0^\Sigma)/M^2}\, ,\nnb
\eea
for $\Xi\Xi\sigma$ and $\Sigma\Sigma\sigma$ couplings, respectively.

We would like to note that the sum rule for $\Xi\Xi\sigma$ coupling constant at the structure $\hat{q}$ is independent of the susceptibilities $\chi$ and $\chi_g$. Another feature of the sum rules above is that up to the dimension we consider, the terms involving the $s$-quark mass do not contribute to the OPE side. The contributions that come from the excited baryon states and the response of the continuum threshold are taken into account by the second and the third terms on the right-hand side (RHS) of the sum rules, respectively.

For the sake of completeness, here we also give the sum rule for $N\!N\sigma$ coupling constant at the structure $\hat{q}$~\cite{Erk05}:
\bea\label{Nsumq}
   &\Big[&-M^4\, a_q\, E^N_0+\frac{4}{3}\, \chi\, M^2\,
   a_q^2\,L^{4/9}-\frac{m_0^2}{2}\, M^2\,
   a_q\,L^{-14/27}-(\chi+\chi_G)\,\frac{m_0^2}{6}\,a_q^2\, L^{-2/27}
\nnb\\&&+2
\,m_q\, M^6 \,E^N_1\,L^{-8/9}+2\,\chi\, m_q \,a_q \,M^4\,
E^N_0\,L^{-4/9}+\frac{2}{3} m_q\,a_q^2
   \Big]~e^{M_N^2/M^2} \nnb \\ &&
   ~~~~~~=-\tilde{\lambda}_N^2\, \frac{M_N}{g_q^\sigma}\, g_{N\!N\sigma}
   + \tilde{B}_N\, \frac{M^2}{g_q^\sigma}+\frac{(s^N_0)^2}{2g_q^\sigma}
   \,\Delta s^N_0\, M^2\,L^{-4/9} e^{(M_N^2-s_0^N)/M^2}\, ,
\eea
which follows from a choice of the interpolating field~\cite{Iof81}
\bea\label{Nintfi}
   \eta_N = \epsilon_{abc}[u_a^T C\gamma_\mu u_b]\gamma_5\gamma^\mu d_c \,\, .
\eea
Note that, in Eq.~(\ref{Nsumq}), the $N\!N\sigma$ sum rule in Ref.~\cite{Erk05} has been improved including the quark mass terms.

Comparing the left-hand sides (LHS) of the sum rules in Eq.~(\ref{Csumq}), Eq.~(\ref{Ssumq}) and Eq.~(\ref{Nsumq}) one can derive a basic relation between the $\Xi\Xi\sigma$, $\Sigma\Sigma\sigma$ and $N\!N\sigma$ coupling constants in the $SU(3)$ limit, which is
\bea\label{couprel} g_{N\!N\sigma}=g_{\Xi\Xi\sigma}+g_{\Sigma\Sigma\sigma}\,.\eea
This relation is quite natural because for the $\Xi\Xi\sigma$ coupling only the $u$-quark propagator outside of the trace and for the $\Sigma\Sigma\sigma$ coupling the $u$-quark propagators inside the trace involve the terms that are proportional to the external field. For the $N\!N\sigma$ coupling all the three quark propagators involve such terms and this implies that in the $SU(3)$ and isospin symmetric limit, the relation in Eq.~(\ref{couprel}) holds. It is interesting to note that this relation can also be derived from Eq.~(\ref{relSU3}) and Eq.~(\ref{Noctsing}) assuming the ideal mixing for the $q\bar{q}$ picture where the $N\!Nf_0$ coupling vanishes.

\section{Analysis of the Sum Rules and Discussion}\label{secAN}
In this section we analyze the sum rules derived in the previous section in order to determine the values of the $\Lambda\Lambda\sigma$, $\Xi\Xi\sigma$ and $\Sigma\Sigma\sigma$ coupling constants. To proceed to the numerical analysis, we arrange the RHS of the sum rules in the form
\bea\label{form}
    f(M^2) = A_B + B_B M^2+C_B M^2 L^{-4/9} e^{(M_B^2-s_0^B)/M^2} \ ,
\eea
and fit the LHS to $f(M^2)$. Here we have defined
\bea\label{form2}
   A_B &\equiv& -\tilde{\lambda}_B^2 \frac{M_B}{g_q^\sigma} g_{B\!B\sigma} \, , \nnb \\
   B_B &\equiv& \frac{\tilde{B}_B}{g_q^\sigma} \, , 
\eea
together with
\bea\label{form2LC} C_\Lambda &\equiv& \frac{\Delta s^\Lambda_0}{2 g_q^\sigma}[(s^\Lambda_0)^2-4\,m_s\,f\,a_q] \, ,\\ C_\Xi &\equiv& \frac{(s^B_0)^2}{2g_q^\sigma} \Delta s^B_0 \, ,\nnb\\
C_\Sigma &\equiv& \frac{\Delta s^\Sigma_0}{2 g_q^\sigma}[(s^\Sigma_0)^2-4\,m_s(f+1)a_q] \, \nnb,\eea
for $\Lambda\Lambda\sigma$,  $\Xi\Xi\sigma$ and $\Sigma\Sigma\sigma$ sum rules, respectively.

For the vacuum parameters, we adopt $a_{q}=0.51\pm 0.03 ~\text{GeV}^3$, and $m_0^2=0.8 ~\text{GeV}^2$ \cite{Ovc88}. We take the renormalization scale $\mu=0.5~\text{GeV}$ and the QCD scale parameter $\Lambda_{QCD}=0.1~\text{GeV}$. The value of the susceptibility $\chi$ has been calculated in Ref.~\cite{Erk05} as $\chi= -10 \pm 1 ~\text{GeV}^{-1}$. The value of the susceptibility $\chi_G$ is less certain. Therefore, we consider $\chi_G$ to change in a wider range.

\subsection{Scalar Meson-Baryon Coupling Constants in the $SU(3)$ Symmetric Limit}
We shall first consider the sum rules in the $SU(3)$ flavor symmetric limit, where we take $m_q=m_s=0$ and $f=0$. In this limit we also set the physical parameters of all the baryons equal to the ones of the nucleon; $M_B = M_N=0.94~\text{GeV}$, $\tilde{\lambda}_B = \tilde{\lambda}_N=2.1 ~\text{GeV}^6$~\cite{Iof84}, $s^B_0 = s^N_0$. 

\begin{figure}[ht]
\includegraphics[scale=0.85]{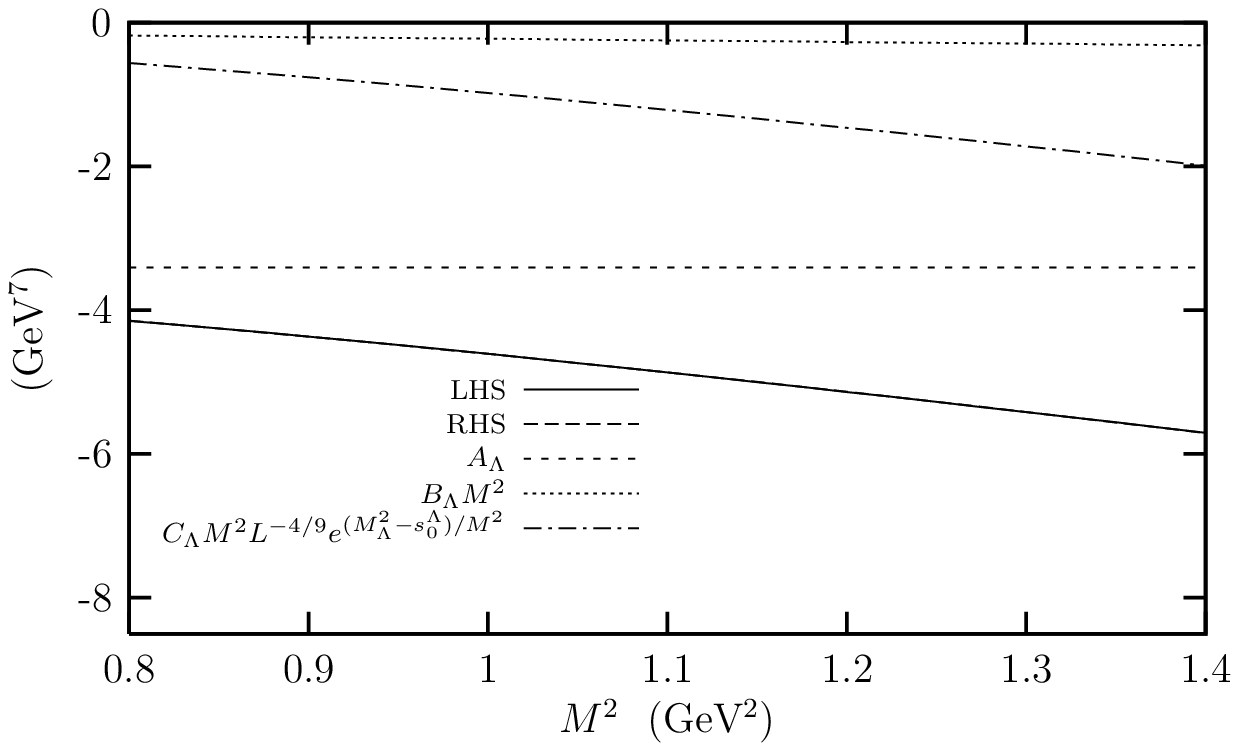}
\caption{The Borel mass dependence of LHS and the fitted RHS of the sum rule for $\Lambda\Lambda              	\sigma$ coupling in Eq.~(\ref{sumq}) for $s^\Lambda_0=2.3$ GeV$^2$ and $\chi_G\equiv\chi=-10		$ GeV$^{-1}$. We also present the terms on the RHS individually.
         Note that the LHS curve (solid) overlies the RHS curve (dashed).} \label{LambdaM}\end{figure}

\begin{figure}[ht]
\includegraphics[scale=0.85]{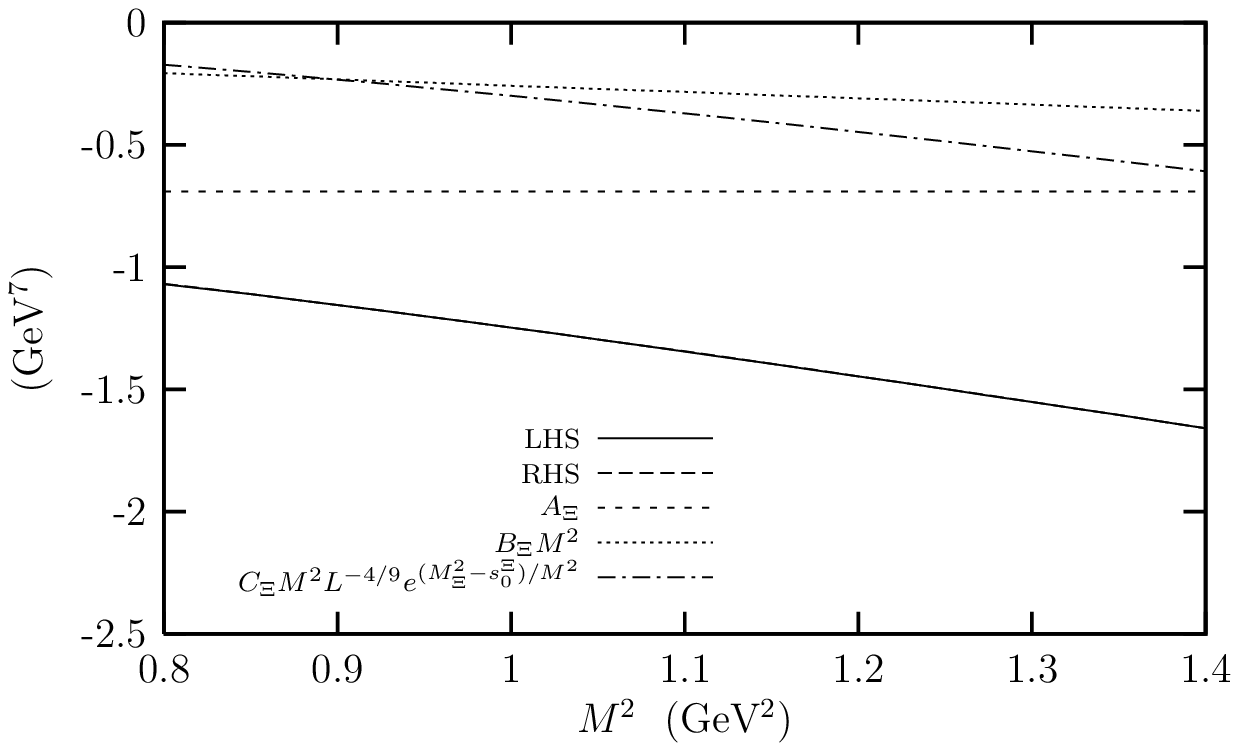}
\caption{Same as Fig.~\ref{LambdaM} but for the sum rule for $\Xi\Xi\sigma$ coupling in 	Eq.~(\ref{Csumq}).} \label{CascadeM}\end{figure}

\begin{figure}[ht]
\includegraphics[scale=0.85]{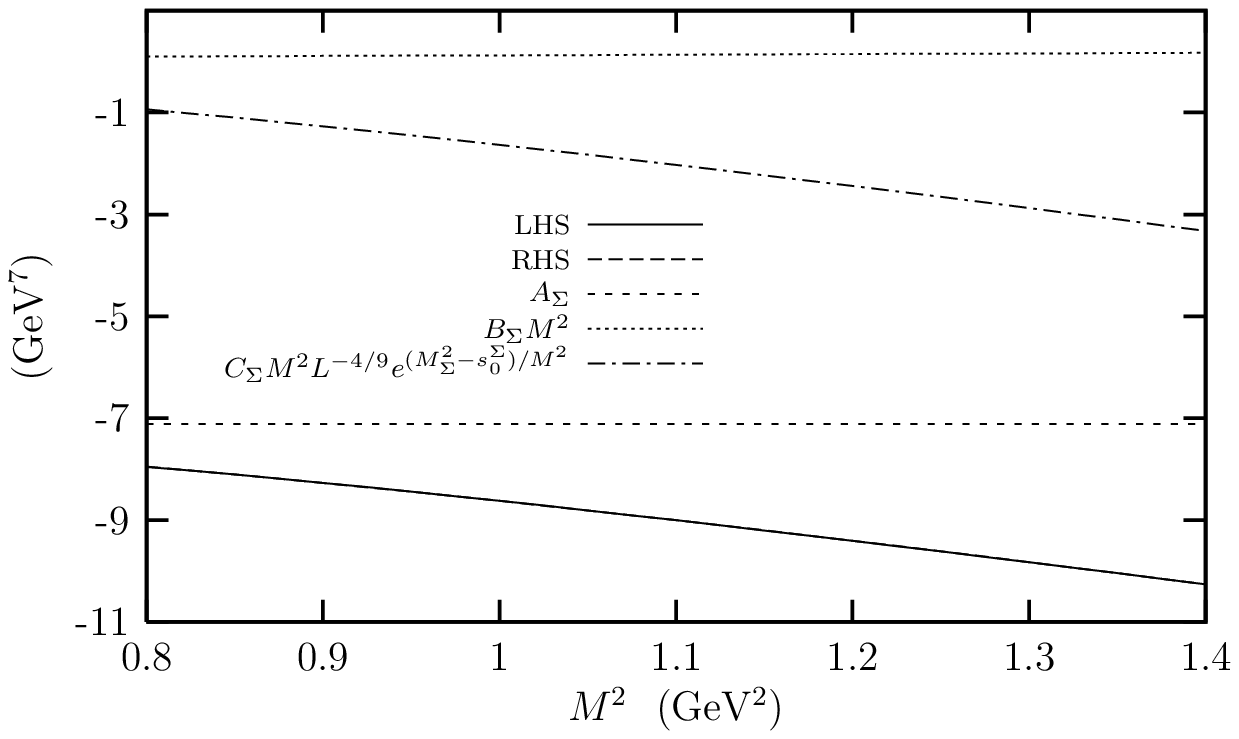}
\caption{Same as Fig.~\ref{LambdaM} but for the sum rule for $\Sigma\Sigma\sigma$ coupling in Eq.~		(\ref{Ssumq}).} \label{SigmaM}\end{figure}

In Figs.~\ref{LambdaM}-\ref{SigmaM} we present the Borel mass dependence of the LHS and the RHS of the sum rules for $\Lambda\Lambda\sigma$, $\Sigma\Sigma\sigma$ and $\Xi\Xi\sigma$, respectively, for $s^B_0=2.3$ and $\chi_G \equiv\chi=-10$ GeV$^{-1}$. As stressed above, in the $SU(3)$ limit we choose the Borel window 0.8 GeV$^2$ $\leq M^2\leq 1.4$ GeV$^2$ which is commonly identified as the fiducial region for the nucleon mass sum rules~\cite{Iof84}. It is seen from these figures that the LHS curves (solid) overlie the RHS curves (dashed). In order to estimate the contributions that come from the excited baryon states and the responses of the continuum threshold, we plot each term on the RHS individually. We observe that the single-pole terms (dotted) give very small contributions to the sum rules except to the one for $\Xi\Xi\sigma$ coupling. The responses of the continuum thresholds (dot-dashed) for all the couplings are quite sizable.

\begin{figure}[ht]
\includegraphics[scale=0.80]{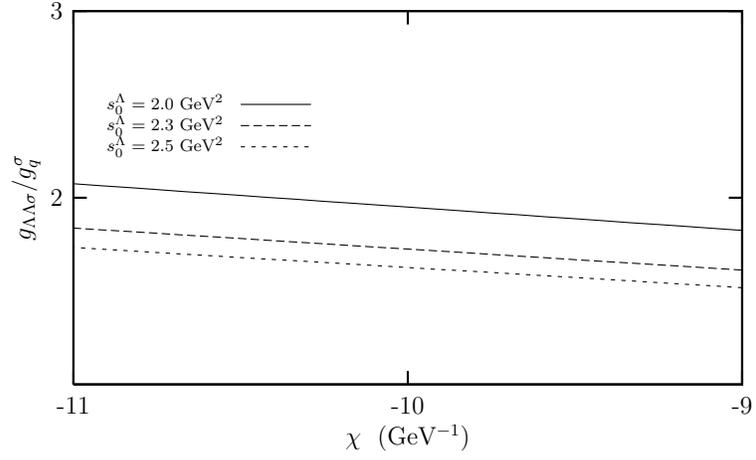}
\caption{The dependence of $g_{\Lambda\Lambda\sigma}/g_q^\sigma$ on the susceptibility
         $\chi$ for three different values of $s_0^q=2.0$, 2.3, and 2.5
        GeV$^2$; here we take $\chi\equiv\chi_G$.} \label{Lambdaks}\end{figure}

\begin{figure}[ht]
\includegraphics[scale=0.80]{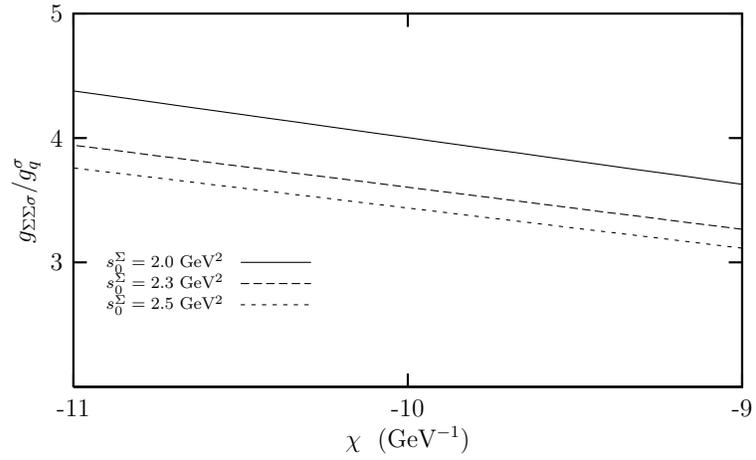}
\caption{Same as Fig.~\ref{Lambdaks} but for the sum rule for the $g_{\Sigma\Sigma\sigma}/g_q^		\sigma$ coupling constant.} 		\label{Sigmaks}\end{figure}

In order to see the sensitivity of the coupling constants on the continuum threshold and the susceptibility $\chi$, we plot in Figs.~\ref{Lambdaks} and \ref{Sigmaks} the dependence of $g_{\Lambda\Lambda\sigma}/g_q^\sigma$ and $g_{\Sigma\Sigma\sigma}/g_q^\sigma$ on $\chi$ for three different values of the continuum thresholds, $s^B_0=2.0$, 2.3, and 2.5 GeV$^2$, and taking $\chi\equiv\chi_G$. One sees that these coupling constants  change by approximately $10\%$ in the considered region of the susceptibility $\chi$. The values of the coupling constants are not very sensitive to a change in the continuum threshold, which gives an uncertainty of approximately $7\%$ to the final values. 

Taking into account the uncertainties in $\chi$, $s_0^B$, and $a_q$, the predicted values for $N\!N\sigma$, $\Lambda\Lambda\sigma$, $\Xi\Xi\sigma$ and $\Sigma\Sigma\sigma$ coupling constants in terms of quark-$\sigma$ coupling constant read 
\footnote{We refer the reader to Ref.~\cite{Erk05} for a detailed analysis of $N\!N\sigma$ coupling constant in QCDSR.}:
\bea \label{mbSU3} g_{N\!N\sigma}/g_q^\sigma= 3.9 \pm 1.0\,, g_{\Lambda\Lambda\sigma}/g_q^\sigma= 1.9 \pm 0.5\,, ~g_{\Xi\Xi\sigma}/g_q^\sigma= 0.4 \pm 0.1\,, ~g_{\Sigma\Sigma\sigma}/g_q^\sigma= 3.8 \pm 1.0\,.\eea
To determine the coupling constants, one next has to assume some value for the quark-$\sigma$ coupling constant $g_q^\sigma$. Adopting the value $g_q^\sigma=3.7$ as estimated from the sigma model~\cite{Ris99}, we obtain
\bea\label{ccSU3}
   g_{N\!N\sigma}= 14.4 \pm 3.7\,,~~g_{\Lambda\Lambda\sigma}=7.0 \pm 1.9\,, ~~g_{\Xi\Xi\sigma}= 1.5 \pm 0.4\,, ~~g_{\Sigma\Sigma\sigma}= 14.1 \pm 3.7\,.
\eea

Note that, the coupling constants in Eq.~(\ref{ccSU3}) are defined at $t=0$, {\em i.e.}
$g_{B\!B\sigma} \equiv g_{B\!B\sigma}(t=0)$. As stressed above, the value of the susceptibility $\chi_G$ is less certain than the value of $\chi$. If we let $\chi_G$ change in a wider range, say $6 ~\text{GeV}^{-1} \leq -\chi_G \leq 14 ~\text{GeV}^{-1}$, this brings an additional $15 \%$ uncertainty to the values of $\Lambda\Lambda\sigma$ and $\Sigma\Sigma\sigma$ coupling constants but the $\Xi\Xi\sigma$ coupling constant remains intact because the sum rule is independent of the susceptibilities as stressed above. In order to keep consistent with the analysis in Ref.~\cite{Erk05}, here we also have taken $\Lambda_{QCD}=0.1~\text{GeV}$. A change in the value of this parameter, say an increase to $\Lambda_{QCD}=0.2~\text{GeV}$, does not have any considerable effect on $\Xi\Xi\sigma$ coupling constant, but the $N\!N\sigma$ and $\Sigma\Sigma\sigma$ coupling constants are increased by approximately $8\%$, while the increase in the value of $\Lambda\Lambda\sigma$ coupling constant is by $5\%$.

Our next concern is to investigate the $SU(3)$ relations for the scalar meson-baryon interactions and see if the coupling constants above as obtained from QCDSR are consistent with these relations. The values of three coupling constants as determined from QCDSR together with the first equation in Eq.~(\ref{Noctsing}) are sufficient to determine the three parameters of flavor $SU(3)$ structure of scalar meson-baryon couplings; namely $g_1$, $g$, and $\alpha_s$. For this purpose, we calculate the coupling constants in Eq.~(\ref{ccSU3}) with the average values of the parameters; $\chi\equiv\chi_g=-10~\text{GeV}^{-1}$, $a_q=0.51~\text{GeV}^2$ and $s^B_0=2.3~\text{GeV}^2$ where we obtain:
\bea \label{mbavSU3} g_{N\!N\sigma}/g_q^\sigma= 4.0,~~~~~~g_{\Lambda\Lambda\sigma}/g_q^\sigma= 1.7\,, ~~~~~~g_{\Xi\Xi\sigma}/g_q^\sigma= 0.3\,, ~~~~~~g_{\Sigma\Sigma\sigma}/g_q^\sigma= 3.6\,.\eea  We first assume $q\bar{q}$ structure with the ideal mixing angle $\theta_s\simeq 35.3^\circ$, and use $g_{N\!N\sigma}$, $g_{\Xi\!\Xi\sigma}$ and $g_{\Sigma\!\Sigma\sigma}$ in Eq.~(\ref{mbavSU3}). The $F/(F+D)$ ratio, $\alpha_s$, can directly be calculated via the relation, \bea \label{relalpha} (g_{\Xi\Xi\sigma}-g_{N\!N\sigma})/(g_{\Sigma\Sigma\sigma}-g_{N\!N\sigma})=\frac{-2\alpha_s}{1-2\alpha_s}\,.\eea  With straightforward algebra, the values of the $F/(F+D)$ ratio, and the octet and the singlet couplings for the $q\bar{q}$ picture are determined as, 
\bea\label{cossu3}\alpha_s=F/(F+D)&=&0.55\,,~~~~ g/g_q^\sigma=g_{N\!Na_0}/g_q^\sigma=\,3.3\,,~~~~g_1/g_q^\sigma=3.2\,.\eea

Inserting $\alpha_s$ and $g$ into the $SU(3)$ relations in Eq.~(\ref{relSU3}) and using the mixing scheme for the singlet and the octet couplings as in Eq.~(\ref{Noctsing}) with the value of $g_1$ in Eq.~(\ref{cossu3}), we observe that the coupling constants as determined from QCDSR in Eq.~(\ref{mbavSU3}) are consistent with the $SU(3)$ relations. This also gives $g_{N\!Nf_0}=0$ with the second equation in Eq.~(\ref{Noctsing}), which is justified by the non-strange content of the nucleon and by the ideal mixing scheme. In Table~\ref{mbcq2} we give all the scalar meson-baryon coupling constants, obtained from these relations, assuming $g_q^\sigma=3.7$.

\begin{table}[t]
\caption{The scalar meson-baryon coupling constants in the $SU(3)$ limit where the $q\bar{q}$ picture for the scalar mesons with the ideal mixing is assumed.}
\centering
\begin{ruledtabular}
\begin{tabular}{crrrrrrr}
M & $N\!NM$ & $\Lambda\Lambda M$ & $\Xi\Xi M$ & $\Sigma \Sigma M$ & $\Lambda\Sigma M$ & $\Sigma N M$ & $\Lambda N M$\\
\hline
$\sigma$ & 14.6 & 6.2 & 1.3 &13.3 & & &  \\
$f_0$ &0 &$-12.0$ &$-18.9$ &$-1.8$ & & &  \\
$a_0$ & 12.0& &1.3 & 13.3 & 6.2 &  &  \\
$\kappa$ & & & & &  &$-1.3$ & $-14.7$
\end{tabular}
\end{ruledtabular}
\label{mbcq2}
\end{table}

In the case of the $q^2\bar{q}^2$ picture, we have a quite distinct ideal mixing scheme for the scalar mesons from that for the $q\bar{q}$ picture. The ideal mixing angle corresponds to $\theta_s\simeq-54.8^\circ$ in this case. In this picture, we assume that the $u$- or the $d$-quark in the baryon couples to the $q\bar{q}$ component in the scalar meson but not to the $s\bar{s}$ component and the $\qqbar$ condensates are modified in the same way. Accordingly, the $s$-quark only couples to the $s\bar{s}$ component. Applying the same procedure as in the $q\bar{q}$ case with $\theta_s=-54.8^\circ$, we obtain the values of $F/(F+D)$ ratio, and the octet and the singlet couplings for the $q^2\bar{q}^2$ picture as:
\bea\label{alpnnaq4}\alpha_s=F/(F+D)&=&0.55\,,~~~~ g/g_q^\sigma=g_{N\!Na_0}/g_q^\sigma=\,-2.3\,,~~~~g_1/g_q^\sigma=4.6\,.\eea Inserting these values into the $SU(3)$ relations in Eq.~(\ref{relSU3}), we observe that the scalar meson-baryon coupling constants as found from QCDSR are consistent with the $SU(3)$ relations, as in the $q\bar{q}$ picture.

In Table~\ref{mbcq4}, we present the scalar meson-baryon coupling constants $g_1$, $g_8$ and $\alpha_s$ in the $q^2\bar{q}^2$ picture, assuming $g_q^\sigma=3.7$. Comparing what we have found for the scalar meson-baryon couplings in the two pictures, the value of the $F/(F+D)$ ratio remains intact, as apparent from Eq.~(\ref{relalpha}), however, the values of $g$ and $g_1$ in the two pictures are quite different from each other. On the other hand, the $B\!Bf_0$ couplings differ very much with regard to the structure of the scalar mesons. In the $q^2\bar{q}^2$ picture for the scalar mesons, in contrary to the $q\bar{q}$ picture, $f_0$ strongly couples to the nucleon due to $\bar{u}u$ and $\bar{d}d$ components it has. The strengths of the $I=1$ couplings in the two pictures differ by a factor of $\sqrt{2}$. 

\begin{table}[t]
\caption{Same as Table \ref{mbcq2} but for the $q^2\bar{q}^2$ picture for the scalar mesons.}
\centering
\begin{ruledtabular}
\begin{tabular}{crrrrrrr}
M & $N\!NM$ & $\Lambda\Lambda M$ & $\Xi\Xi M$ & $\Sigma \Sigma M$ & $\Lambda\Sigma M$ & $\Sigma N M$ & $\Lambda N M$\\
\hline
$\sigma$ & 14.6 & 6.2 & 1.3 &13.3 & & &  \\
$f_0$ & 10.3 & 16.2 & 19.7 & 11.3 & & &  \\
$a_0$ & $-8.5$& &$-0.9$ & $-9.4$ & $-4.4$ &  & \\ 
$\kappa$ & & & &  &  & 0.9 & 10.3
\end{tabular}
\end{ruledtabular}
\label{mbcq4}
\end{table}

\subsection{Sigma Meson-Baryon Coupling Constants with $SU(3)$ Breaking Effects}
Now we turn to the effect of $SU(3)$ breaking, where we allow $m_s=0.15~\text{GeV}$ and $f=-0.2$, keeping $m_q=0$. We also restore the physical values for the parameters of baryons~\cite{Hwa94,Chi85}: \bea\label{parSU3b} M_\Lambda&=&1.1~\text{GeV}\,,~~~~M_\Xi=1.3~\text{GeV}\,,~~~~M_\Sigma=1.2~\text{GeV}\,, \\ \tilde{\lambda}_\Lambda^2&=&3.3 ~\text{GeV}^6\,,~~~~ \tilde{\lambda}^2_\Xi=4.6 ~\text{GeV}^6\,,~~~~ \tilde{\lambda}^2_\Sigma=3.3 ~\text{GeV}^6\,,\nnb\\ s_0^\Lambda&=& 3.1 \pm 0.3~\text{GeV}^2\,,~~~~s_0^\Xi= 3.6 \pm 0.4~\text{GeV}^2\,,~~~~s_0^\Sigma= 3.2 \pm 0.3~\text{GeV}^2\,.\nnb\eea The corresponding Borel windows are chosen as: \bea\label{bwSU3b}\text{for }\Lambda\,,~~~~1.0~\text{GeV}^2 \leq M^2 \leq 1.4~\text{GeV}^2\,,\\ \text{for }\Xi\,,~~~~1.5~\text{GeV}^2 \leq M^2 \leq 1.9~\text{GeV}^2\,,\nnb\\ \text{for }\Sigma\,,~~~~1.2~\text{GeV}^2 \leq M^2 \leq 1.6~\text{GeV}^2\,,\nnb\eea

\begin{figure}[ht]
\includegraphics[scale=0.80]{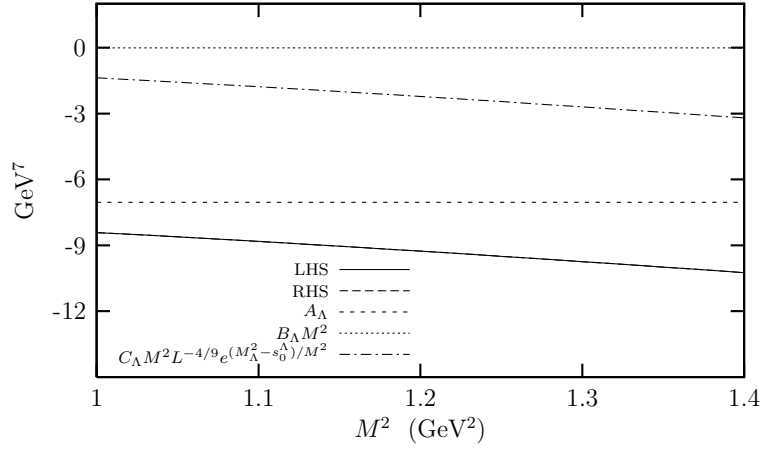}
\caption{The Borel mass dependence of LHS and the fitted RHS of the sum rule for $\Lambda\Lambda              	\sigma$ coupling in Eq.~(\ref{sumq}) with the $SU(3)$ breaking effects and for $\chi_G\equiv			\chi=-10$ GeV$^{-1}$. We also present the terms on the RHS individually.
         Note that the LHS curve (solid) overlies the RHS curve (dashed).} \label{LambdaSU3b}\end{figure}

\begin{figure}[ht]
\includegraphics[scale=0.85]{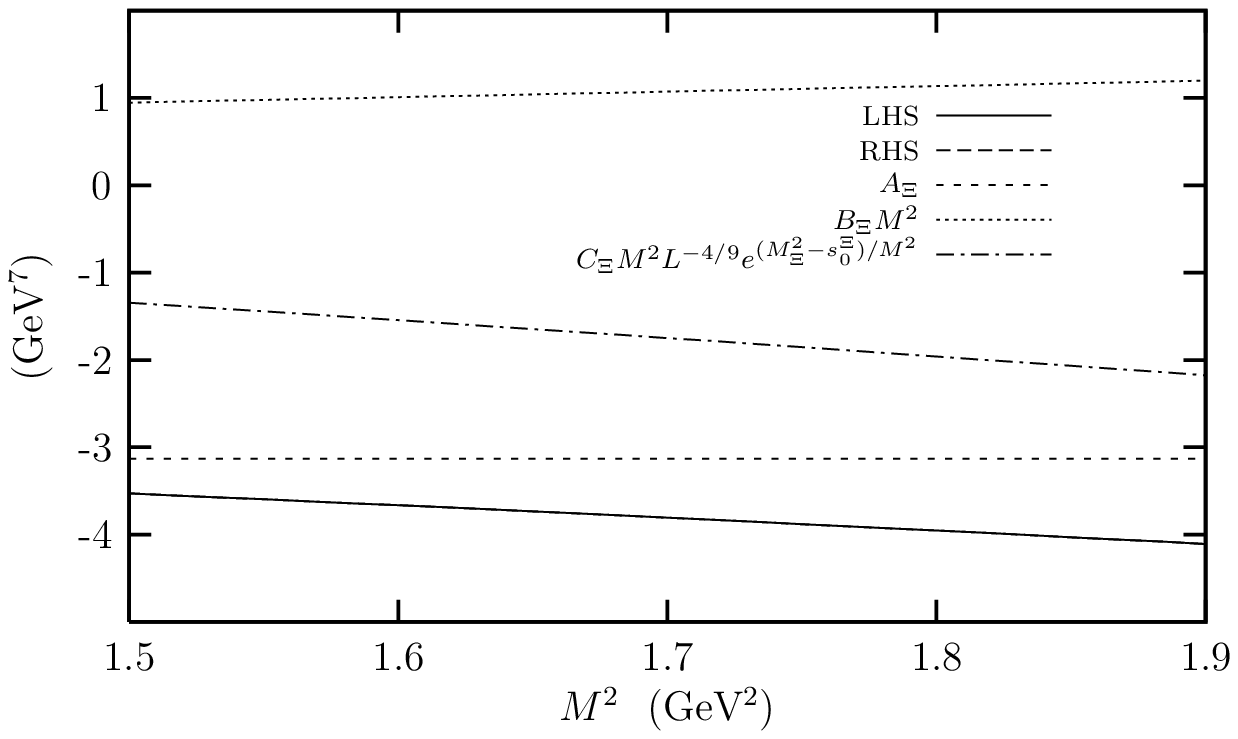}
\caption{Same as Fig.~\ref{LambdaSU3b} but for the sum rule for $\Xi\Xi\sigma$ coupling in Eq.~(\ref{Csumq}).} \label{CascadeSU3b}\end{figure}

\begin{figure}[ht]
\includegraphics[scale=0.85]{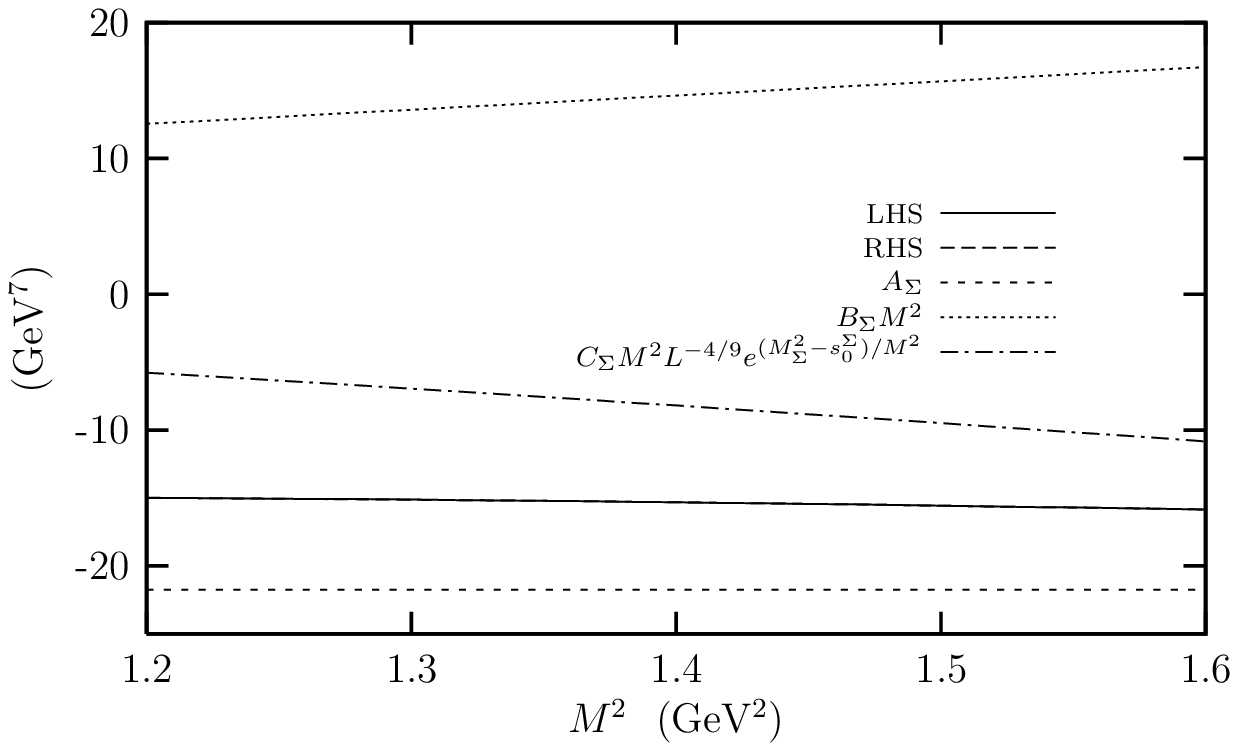}
\caption{Same as Fig.~\ref{LambdaSU3b} but for the sum rule for $\Sigma\Sigma\sigma$ coupling in   		Eq.~	(\ref{Ssumq}).} \label{SigmaSU3b}\end{figure}

In Figs.~\ref{LambdaSU3b}-\ref{SigmaSU3b} we present the Borel mass dependence of the LHS and the RHS of the sum rules for $\Lambda\Lambda\sigma$, $\Sigma\Sigma\sigma$ and $\Xi\Xi\sigma$, respectively, with the $SU(3)$ breaking effects. We plot each term on the RHS individually as we did in the $SU(3)$ limit. We observe that the responses of the continuum thresholds for all the couplings are quite sizable. The contributions of the single pole terms to the $\Xi\Xi\sigma$ and $\Sigma\Sigma\sigma$ sum rules are large and opposite in sign to the third terms on the RHS. Therefore the contributions of these two terms tend to cancel each other which leads to a very stable sum rule for these couplings. The contribution of the single pole term for the $\Lambda\Lambda\sigma$ coupling is very small.

Taking into account the uncertainties in $\chi$, $s_0^B$, and $a_q$, the predicted values for $\Lambda\Lambda\sigma$, $\Xi\Xi\sigma$ and $\Sigma\Sigma\sigma$ coupling constants in terms of quark-$\sigma$ coupling constant with the $SU(3)$ breaking effects read:
\bea \label{mbSU3b} g_{\Lambda\Lambda\sigma}/g_q^\sigma= 2.0 \pm 0.5\,, ~~~~~~g_{\Xi\Xi\sigma}/g_q^\sigma= 0.5 \pm 0.1\,, ~~~~~~g_{\Sigma\Sigma\sigma}/g_q^\sigma= 5.7 \pm 1.4\,.\eea
Adopting again the value $g_q^\sigma=3.7$ as estimated from the sigma model~\cite{Ris99}, we obtain
\bea\label{ccSU3b}
  g_{\Lambda\Lambda\sigma}= 7.4 \pm 1.9\,, ~~~~~~g_{\Xi\Xi\sigma}= 1.9 \pm 0.4\,, ~~~~~~g_{\Sigma\Sigma\sigma}= 21.1 \pm 5.2\, .
\eea

A few remarks are in order now. Comparing the values obtained from the sum rules in the $SU(3)$ symmetric limit and the ones beyond the $SU(3)$-limit, we observe that the introduction of the $SU(3)$ breaking effects does not change the $\Lambda\Lambda\sigma$ coupling constant, while the $\Xi\Xi\sigma$ and $\Sigma\Sigma\sigma$ couplings are modified by approximately $\%30$ and $\%50$, respectively. We also note that, the obtained value of $\Lambda\Lambda\sigma$ coupling constant is small as compared to the $N\!N\pi$ and $N\!N\sigma$ coupling constants. Since the $\sigma$ exchange gives the dominant contribution in the $\Lambda\Lambda$ system, this suggests that the $\Lambda\Lambda$ interaction is weak, in accordance with the recent experimental result. 

\section{Discussion and Conclusions}\label{secCONC}
We have calculated the $\Lambda\Lambda\sigma$, $\Xi\Xi\sigma$ and $\Sigma\Sigma\sigma$ coupling constants which play significant roles in OBE models of YN and YY interactions, employing the external field QCDSR method. The coupling constants can be determined in terms of quark-$\sigma$ coupling constant in this method. In order to compare our results with the others in the literature and keep as model-independent as possible, we find it useful to give the ratios of the coupling constants in the $SU(3)$ limit for the average values of the vacuum parameters,
\bea\label{avmbdis}\frac{g_{\Lambda\Lambda\sigma}}{g_{N\!N\sigma}}=0.43,~~~~\frac{g_{\Xi\Xi\sigma}}{g_{N\!N\sigma}}=0.08,~~~~\frac{g_{\Sigma\Sigma\sigma}}{g_{N\!N\sigma}}=0.91\,.\eea
We observe that the $\Xi\Xi\sigma$ coupling constant is more than one order small as compared to $N\!N\sigma$ coupling constant. The $\Sigma\Sigma\sigma$ coupling constant is at the same order with $N\!N\sigma$ coupling constant and twice as large as the $\Lambda\Lambda\sigma$ coupling constant. We have shown that these coupling constants as determined from QCDSR satisfy the $SU(3)$ relations which lead to a determination of the $F/(F+D)$ ratio for the scalar octet. Although the scalar meson-baryon coupling constants depend on the picture assumed for the structure of the scalar mesons ($q\bar{q}$ or $q^2\bar{q}^2$), the $F/(F+D)$ ratio remains intact in the two pictures. We would like to also note that the third terms on the RHS's of the sum rules which represent the responses of the continuum thresholds, affect only the values of the individual coupling constants, which receive contribution by the same factor. Therefore the ratios of the coupling constants and the value of $\alpha_{s}$ remain unchanged if this term is omitted.

All the Nijmegen soft-core OBE potential models have $\theta_s > 30^\circ$, which points to almost ideal mixing angles for the scalar $q\bar{q}$ states. Since the ideal mixing angle for the scalar mesons has been assumed in our QCDSR calculations as well, it is convenient to compare our results with the ones from NSC potential models. Our result for the ratio $g_{\Lambda\Lambda\sigma}/g_{NN\sigma}$ is half of the value found in Ref.~\cite{Fer05}, however, it qualitatively agrees with the one from NSC89~\cite{Mae89}, which is $g_{\Lambda\Lambda\sigma}/g_{NN\sigma}=0.58$. The value we have obtained for the $\Lambda\Lambda\sigma$ coupling constant is small as compared to $N\!N\sigma$ coupling constant and this implies that $\Lambda\Lambda$ interaction is weak, since the sigma exchange gives the dominant contribution to this interaction in terms of OBE models. The value of the $F/(F+D)$ ratio, which is $0.55$ as obtained from QCDSR is about half of the values given in NSCa-f~\cite{Rij98}, which is $F/(F+D)\simeq1.1$\,. 

In order to estimate the $SU(3)$ breaking in the couplings, we have restored the physical values of the parameters like the strange quark mass and the physical baryon masses. We observe that the $SU(3)$ breaking effects do not change the $\Lambda\Lambda\sigma$ coupling, while the $\Xi\Xi\sigma$ and $\Sigma\Sigma\sigma$ couplings are modified largely. It is also desirable to derive the sum rules for the $BBf_0$ and $BBa_0$ couplings in order to estimate the $SU(3)$ breaking in these coupling constants.
 

\newpage
         
\end{document}